\documentclass[aps,prl,twocolumn,superscriptaddress]{revtex4-1}   %
\usepackage{amsmath,amsfonts,amssymb,amsthm,bm}
\usepackage{graphicx}
\usepackage[colorlinks=true,citecolor=blue,breaklinks=true,linkcolor=red,
linktocpage=true,urlcolor=blue,pagebackref=false]{hyperref}
\usepackage[normalem]{ulem}

\newcommand{\tr}{\ensuremath{\operatorname{tr}}}

\newcommand{\sgn}{\ensuremath{\operatorname{sgn}}}
\newcommand{\abs}[1]{\ensuremath{|#1|}}

\newcommand{\braketop}[3]{\ensuremath{\langle #1|#2|#3\rangle}}

\newcommand{\diag}{\ensuremath{\operatorname{diag}}}
\newcommand{\zero}{{\ensuremath{{0}}}}
\newcommand{\id}{{\ensuremath{\openone}}} 

\renewcommand{\paragraph}[1]{{\par\it #1.---}\ignorespaces}

\newcommand{\void}[1]{}
\begin{document}
	
	\title{Phase-controlled spin and charge currents in superconductor-ferromagnet hybrids}
	
	\author{Ali Rezaei}
	\affiliation{Department of Physics, University of Konstanz, D-78457 Konstanz, Germany}
	\author{Robert Hussein}
	\affiliation{Department of Physics, University of Konstanz, D-78457 Konstanz, Germany}
	\author{Akashdeep Kamra}
	\affiliation{Center for Quantum Spintronics, Department of Physics, Norwegian University of Science and Technology, NO-7491 Trondheim, Norway}
	\author{Wolfgang Belzig}
	\email{wolfgang.belzig@uni-konstanz.de}
	\affiliation{Department of Physics, University of Konstanz, D-78457 Konstanz, Germany}
	
	\date{\today}
	
	\begin{abstract}
		We investigate spin-dependent quasiparticle and Cooper-pair transport through a central node interfaced with two superconductors and two ferromagnets.
		We demonstrate that voltage biasing of the ferromagnetic contacts induces superconducting triplet correlations on the node and reverses the supercurrent flowing between the two superconducting contacts. We further find that
		such triplet correlations can mediate a tunable spin current flow into the ferromagnetic contacts. 
		Our key finding is that unequal spin-mixing 
		conductances for the two interfaces with the ferromagnets result in equal-spin triplet correlations on the node, detectable via a net charge current between the two magnets.
		Our proposed device thus enables the generation, control, and detection of the typically elusive equal-spin triplet Cooper pairs.%
	\end{abstract}
	\maketitle

	Cooper-pairs from a superconductor (S) placed in the vicinity of a ferromagnet (F) may diffuse into the latter, thus, modifying their electronic properties~\cite{MeissnerPR1960a,DeGennesRMP1964a,McMillanPR1968a,KlapwijkJS2004a,BuzdinRMP2005a}.
	The engineering of this proximity effect in hybrid structures generated  over the last few years considerable interest in thermoelectricity~\cite{MachonPRL2013a,MachonNJP2014a,OzaetaPRL2014a,GiazottoPRApplied2015a,KolendaPRL2016a,KolendaBeilsteinJN2016a,KolendaPRB2017a,RezaeiNJP2018a,
	SanchezPRB2018a,HusseinPRB2019a,KirsanovPRB2019a}, spin calorics~\cite{LinderPRB2016a,BathenSR2017a}, and
	topological superconductivity~\cite{ColePRB2015a,ChiuPRB2016a,HaltermanPRB2018a,RenNature2019a,BlasiPRB2019a,KiendlArxiv2019a}. 
	Somehow unexpected is that not only ferromagnets~\cite{JohnsonAPL1994a,QuayNP2013a,BergeretRMP2018a,HeikkilaeArxiv2019a}  and antiferromagnetic insulators~\cite{KamraPRL2018a} 
	can cause spin imbalances into superconductors, but also normal metals~\cite{WolfPRB2013a} exploiting that a superconductor itself may serve as a spin filter.
	
	By sandwiching a  normal metal or ferromagnet between two superconductors, one can realize a Josephson junction featuring
	a current of Cooper-pairs between them, which is characterized by the junctions' free energy~\cite{JosephsonAP1965a,BlochPRB1970a,Beenakker1992a}.  
	Its global minimum determines the ground state of the system. 
	While a ground state at zero phase difference indicates a Josephson current mediated by singlet Cooper-pairs, a shifted 
	groundstate about $\pi$---occurring in magnetic 
	Josepshon junctions~\cite{RyazanovPRL2001a,KeizerNature2006a,EschrigNatureP2008a,BobkovaPRB2010a,RobinsonScience2010a,AnwarPRB2010a,KhairePRL2010a,AnwarAPL2012a}---signals 
	triplet superconductivity~\cite{GolubovRMP2004a,KlamPRB2014a,DelagrangePhysicaB2018a} manifesting in a reversed current phase relation (CPR)~\cite{BaselmansSuperlatticesM1999a,KontosPRL2002a,HouzetPRB2007a}.
	Such magnetic Josepshon junctions are interesting for quantum computation~\cite{BlatterPRB2001a,YamashitaPRL2005a} and cryogenic memories~\cite{NiedzielskiSST2015a,GingrichNP2016a}. 
	\begin{figure}[h!]
		\includegraphics[width=0.9\columnwidth]{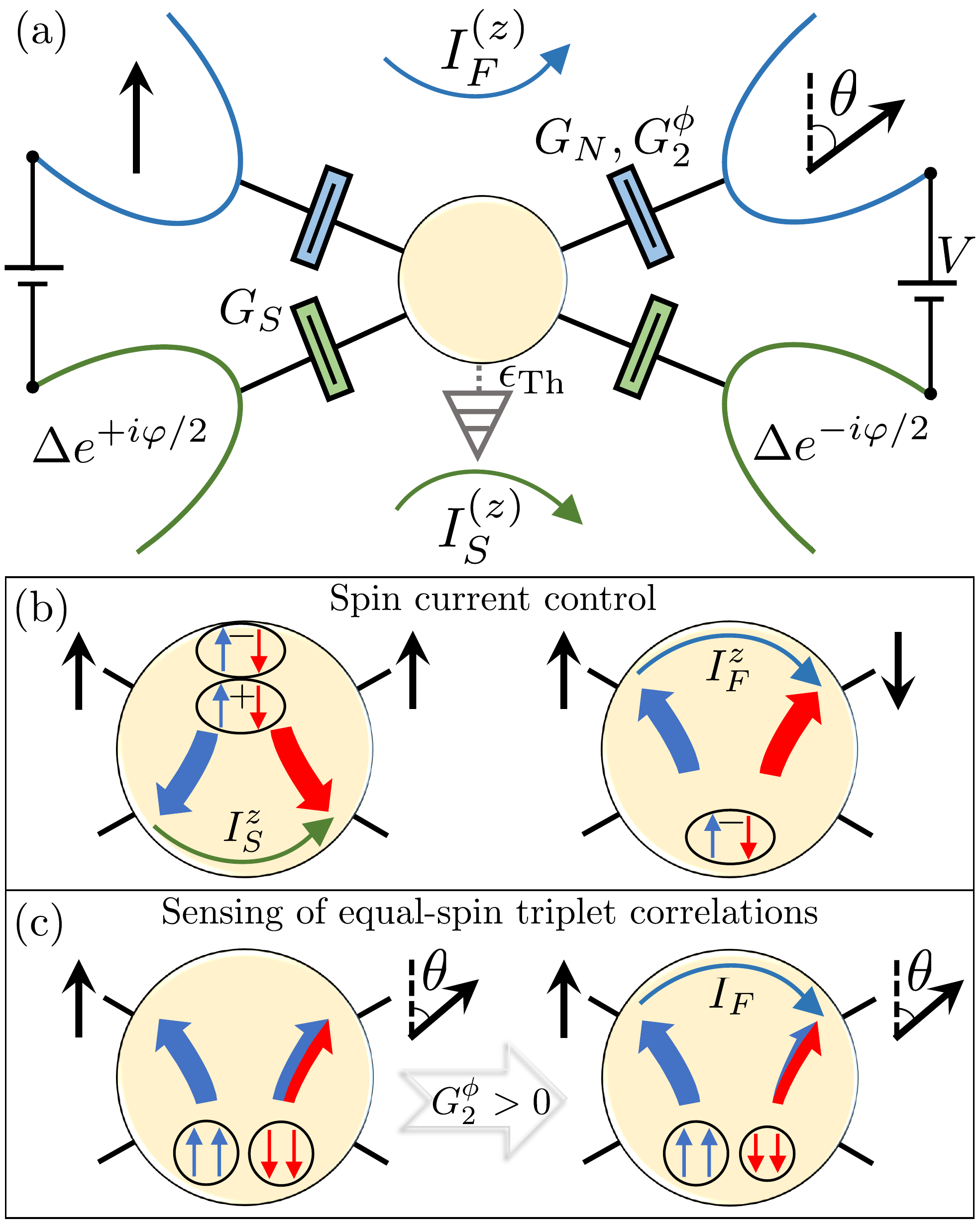}\caption{\label{fig.:sketch} 
		(a) $4$-terminal circuit consisting of two superconductors with phases $\pm\varphi/2$ and gap $\Delta$ (green bordered region), 
		and two ferromagnets with relative magnetization angle $\theta$ (blue bordered region), interfaced with a central node (yellow shaded region).
		(b) The left (right) sketch of the central node illustrates a net spin current injection into the superconductors (ferromagnets) due to (anti)parallel magnetization 
		of the ferromagnetic leads for $\epsilon_{\rm Th}\ll\Delta$ and $V\gtrsim\Delta/e$ in absence of the spin-mixing conductances, $G_1^\phi=G_2^\phi=0$. Singlet (mixed-triplet) correlations are indicated by encircled arrows containing a minus (plus) sign.
		(c) Finite spin-mixing, $G_2^\phi>0$, allows the generation and detection of equal-spin triplet correlations  (encircled arrows of the same color) by measuring a net charge current between the ferromagnetic
		leads.
		}
	\end{figure}
	
	Recently, the generation of equal-spin triplet pairs has been demonstrated in S/F structures utilizing inhomogeneous magnetic fields~\cite{DieschNC2018a}, 
	as originally predicted by Kadigrobov~\cite{KadigrobovEL2001a} and Bergeret~\cite{BergeretPRL2001a,BergeretPRB2001a} et al. Mixed-spin triplet pairs with a zero spin projection on the $z$-axis, however, already arise for homogeneous magnetization.
	It is the immunity of equal-spin triplet correlations against internal magnetic fields which causes long penetration length into ferromagnets compared to the ones of singlet and mixed-triplet correlations~\cite{VolkovPRL2003a,BergeretRMP2005a,BakerNJP2014a}. 
	This property of equal-spin triplet pairs makes them particular attractive for low-power spintronics~\cite{EschrigPToday2011a,LinderNatPhys2015a,EschrigRPP2015a}.

	\begin{figure*}[t]
		\includegraphics[]{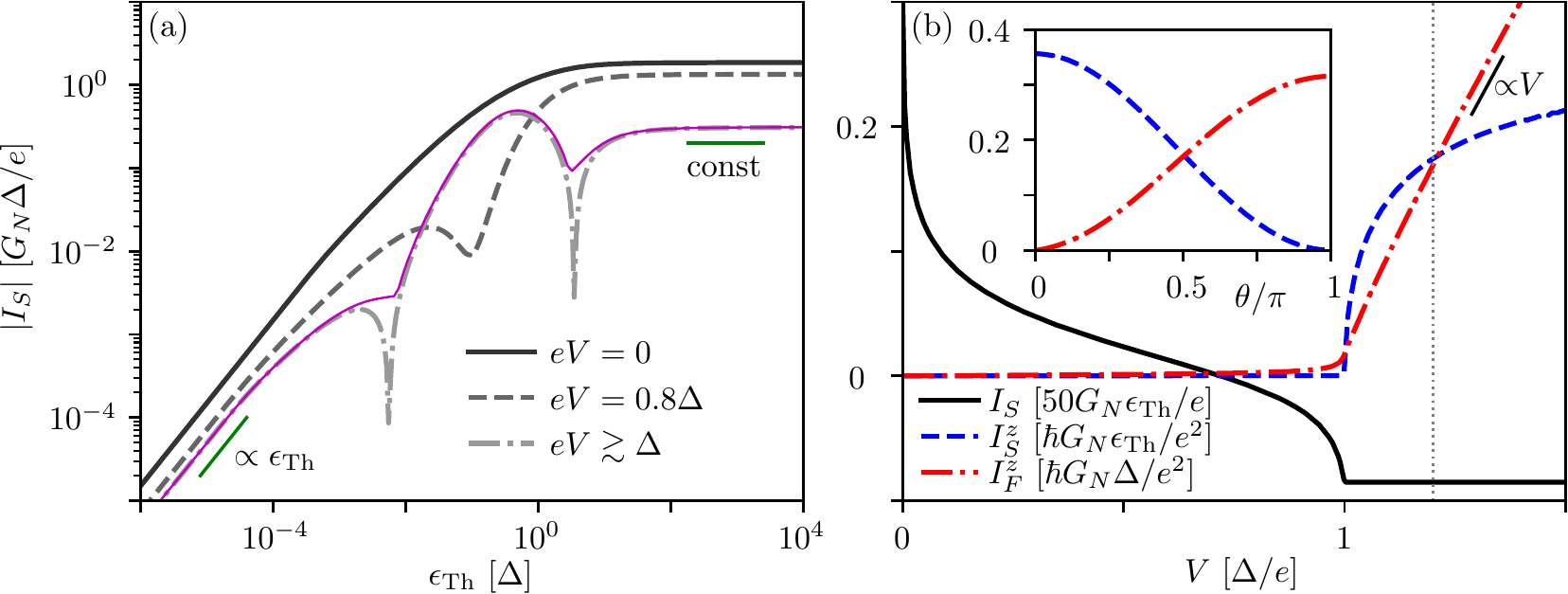}\caption{\label{fig.:largeIslandLimit} 
		(a) Modulus of the net supercurrent $I_S$ as a function of the Thouless energy $\epsilon_{\rm Th}$ for different voltages $V$. 
		The divergencies occurring for large voltages, $V\gtrsim\Delta/e$, indicate $0$--$\pi$-transitions. Parameters are $\theta=0$, $\varphi=\pi/2$, and $G_1^\phi=G_2^\phi=0$.
		The thin purple line indicates the corresponding critical current $I_C=\max_\varphi I_S$ for $V\gtrsim\Delta/e$.
		(b) Net supercurrent $I_S$, and net spin currents $I_S^z$ and $I_F^z$ between the superconducors/ferromagnets for $\theta=\varphi=\pi/2$ and $\epsilon_{\rm Th}\ll\Delta$. The inset shows
		the net spin currents as a function of $\theta$ for $V=1.2\Delta/e$, as indicated by the horizontal dotted line in panel (b).
		}
	\end{figure*}
	In this Letter, we study spin and charge transport in a four-terminal S/F system consisting of two s-wave superconducting ($n{=}S_1,S_2$) and two ferromagnetic ($n{=}F_1,F_2$) terminals with a common contact region, 
	Fig.~\ref{fig.:sketch}(a). Our key finding is that a voltage applied to the two ferromagnetic leads in combination with noncollinear magnetizations can induce triplet superconducting correlations in the system and, therewith, transitions in
	the CPR. We show that the relative magnetization angle $\theta$ qualifies for voltages above the superconducting gap as a control knob for spin currents in the ferromagnetic
	and superconducting contacts [Fig.~\ref{fig.:sketch}(b)]. Finally, we highlight that asymmetric spin-mixing conductances (e.g. $G^\phi_1=0$, $G^\phi_2>0$) can induce for non-collinear magnetization, $0<\theta<\pi$, but otherwise
	symmetric configuration, indeed, a finite net charge current between the ferromagnetic contacts [Fig.~\ref{fig.:sketch}(c)]. 
	This effect is attributed to the generation of equal-spin triplet correlations in the central node, and allows for their experimental detection and exploitation in a convenient manner.

	\paragraph{Method}
	We study diffusive transport, within a semiclassical~\cite{LarkinJETP1969a,ShelankovJLowTP1985a,RammerRMP1986a,LambertJP1998a,BelzigSuperlatticesM1999a}  circuit theory~\cite{NazarovPRL1994a,NazarovSuperlatticesM1999a}. In this framework, hybrid structures are
	discretized as a network of nodes, terminals, and connectors. Here, we map our system to a layout consisting of a central node which is interfaced with two superconducting and two ferromagnetic terminals via corresponding connectors, see Fig.~\ref{fig.:sketch}(a).  
	An additional \textit{leakage} terminal can account for losses of superconducting correlations. 
	The Green function $\check{\mathcal{G}}_c$ of the node, which is an $8\times8$-matrix in Keldysh-Nambu-spin space, is here the central quantity characterizing the transport properties.
	The  conservation of the total matrix current at this node, $\zero_8=\sum_{n}\check{\mathcal{I}}_{n}$, together with the normalization 
	condition $\check{\mathcal{G}}_c^2=\id_{8}$ determines $\check{\mathcal{G}}_c$ and, therewith, the individual matrix currents 
	$\check{\mathcal{I}}_{n}\equiv[\check{\mathcal{M}}_n,\check{\mathcal{G}}_c]$ between terminal $n\in\{S1,S2,F1,F2,\textrm{Leak}\}$ 
	and the central node~\footnote{ The symbol $\zero_n$ ($\id_n$) labels the zero (identity) matrix in $n\times n$ dimensions and the superscript $\check{\ldots}$
	indicates the Keldysh$\otimes$Nambu$\otimes$spin space of dimension $8\times8$.
	}. 
	
	The superconducting contacts are characterized by the BCS bulk Green functions
	$(2/G_S)\check{\mathcal{M}}_{S\alpha}= \check{\mathcal{G}}_{S\alpha}$ with $G_S$ being the conductance of the corresponding connector
	to the central node, and $\alpha=1,2$. Its retarded/advanced component reads in the spinor basis 
	$\{\Psi^{\dagger}_{\uparrow},\Psi^{\dagger}_{\downarrow},\Psi^{}_{\downarrow},-\Psi^{}_{\uparrow}\}$~\cite{MachonPRL2013a,MachonNJP2014a} 
	\begin{equation}
		\hat{\mathcal{G}}_{S\alpha}^{R,A} {=} \frac{\pm\sgn(\epsilon)}{\sqrt{(\epsilon\pm i\Gamma)^2-\abs{\Delta_\alpha}^2}}\!\begin{pmatrix}
			\pm i \Gamma +\epsilon & \Delta_{\alpha} \\
			-\Delta_{\alpha}^* & \mp i \Gamma -\epsilon
		\end{pmatrix}\otimes \id_{2}
		\label{eq.:GSR},
	\end{equation}
	whereby $\Delta_{1,2}\equiv\Delta\exp[\pm i\varphi/2]$  denotes the superconducting gap  with phase difference $\varphi$ across the junction.
	A small imaginary component in the denominator of Eq.~\eqref{eq.:GSR} (here $\Gamma=10^{-3}\Delta$) accounts 
	for a finite lifetime $\hbar/\Gamma$ of the quasiparticles, with energy $\epsilon$, smearing out the superconducting gap~\cite{DynesPRL1984a}.
	The ferromagnetic contacts are governed by
	$\check{\mathcal{M}}_{F\alpha} = (G_N/2)[(\id_8+P\check{\kappa}_{\alpha})\check{\mathcal{G}}_{F}-iG^{\phi}_{\alpha}\check{\kappa}_{\alpha}]$
	with $G_N$ ($G^{\phi}_{\alpha}$) being the normal (spin-mixing) conductances of the corresponding connectors.  $P$ denotes the contact spin polarization, and $\check{\kappa}_{\alpha}=\id_2\otimes{\sigma}_z\otimes(\bm{m}_\alpha\cdot \bm{\sigma})$ is te spin matrix which is diagonal in Keldysh space. 
	In the last expression, $\bm{\sigma}=\{\sigma_x,\sigma_y,\sigma_z\}$ labels the vector of Pauli matrices and $\bm{m}_\alpha$ is the magnetization vector corresponding to the ferromagnet $F\alpha$.
	Here, we fix $\bm{m}_1=(0,0,1)$ in $z$-direction and consider $\bm{m}_2=(\sin\theta,0,\cos\theta)$ tilted by an arbitrary angle $\theta$. We further consider fully polarized ferromagnetic contacts, i.e. $P=1$.
	The retarded/advanced component of the ferromagnetic Green function is given by $\hat{\mathcal{G}}_{F}^{R,A}=\pm\sigma_z\otimes\id_2$.
	Finally, the  leakage  terminal is described by
	$\check{\mathcal{M}}_{\rm{Leak}} = -i G_S(\epsilon/\epsilon_{\rm Th})\id_2\otimes\sigma_z\otimes\id_2$ with $\epsilon_{\rm Th}$ being the Thouless energy.
	The Keldysh component of the S and F Green functions follows from $\hat{\mathcal{G}}_{n}^{K}=\hat{\mathcal{G}}_{n}^{R}{\hat h}_n-{\hat h}_n\hat{\mathcal{G}}_{n}^{A}$ with the distribution function ${\hat h}_n=\diag\bm{(}\tanh[(\epsilon-eV_n)/2k_BT],\tanh[(\epsilon+eV_n)/2k_BT]\bm{)}\otimes\id_2$. 
	Hereafter, we assume all contacts at zero temperature, $k_BT =0$, the superconductors at zero (reference) voltage, $V_{S1}=V_{S2}=0$,
	equally biased ferromagnetic contacts, $V\equiv V_{F1}= V_{F2}$, and equal conductances $G_S=G_N$.
	
	The Keldysh component of the matrix currents $\check{\mathcal{I}}_{n}$ leads to the charge currents
	\begin{align}
	I_n &= \frac{1}{8e}\int_{-\infty}^{\infty}d\epsilon
\tr[(\sigma_z\otimes\id_2)\hat{I}^K_{n}(\epsilon)]
\label{eq.chargecurrents},
	\end{align}
	and the $z$-polarized spin currents 
	\begin{align} 
I_{n}^{z} &= \frac{\hbar}{16e^2}\int_{-\infty}^{\infty}d\epsilon
\tr[(\id_2\otimes \sigma_z)\hat{I}^K_{n}(\epsilon)]
\label{eq.:spincurrents}
	\end{align}
	\cite{NazarovSuperlatticesM1999a,MortenPRB2008a,AlidoustPRB2010a,HaltermanSST2016a}.
	In particular, we will analyze the charge, $I_{X}=I_{X1}-I_{X2}$, and spin net current, $I_X^{z}=I_{X1}^{z}-I_{X2}^{z}$ between the superconductors/ferromagnets ($X=S, F$). 
	The matrix elements $f_{ss'}=\braketop{\Psi_{s}}{\hat{\mathcal{G}}_c^K}{\Psi_{s'}}$
	with $s,s'=\uparrow,\downarrow$ contain the spectral information about spin-pair correlations in nonequilibrium. Here, we consider the integrated quantities
	over positive energies $\epsilon>0$ (triplet correlations are odd in $\epsilon$),  to quantify singlet,  $F_{S}=\int d\epsilon\;(f_{\uparrow\downarrow}-f_{\downarrow\uparrow})/\sqrt{2}$, mixed-spin triplet,
	$F_{T0}=\int d\epsilon\;(f_{\uparrow\downarrow}+f_{\downarrow\uparrow})/\sqrt{2}$, and equal-spin triplet correlations $F_{Ts}=\int d\epsilon\;f_{ss}$.

	\begin{figure}[t]
		\includegraphics[]{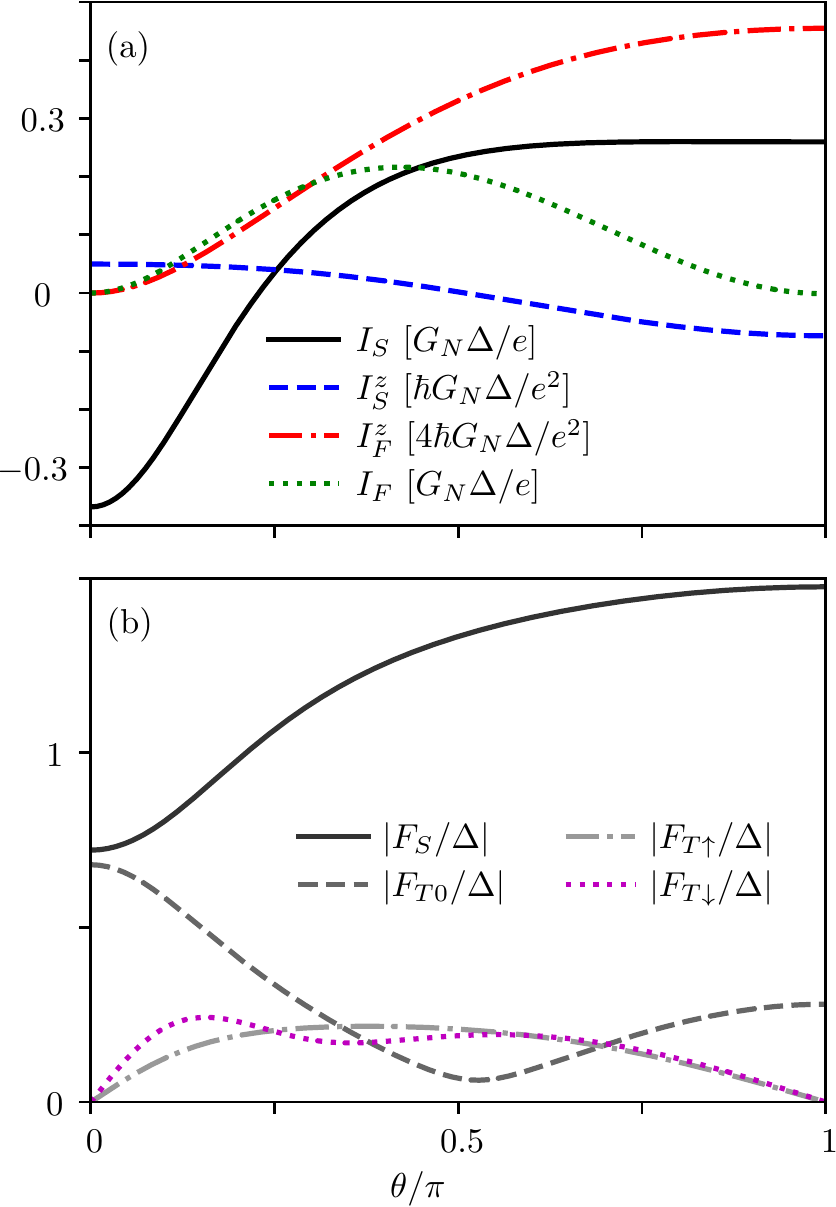}\caption{\label{fig.:finiteSpinMixing} %
		(a) Net charge $I_S$, $I_F$, and net spin currents $I_S^z$,  $I_F^z$, as a function of $\theta$ for $\epsilon_{\rm Th} \gg \Delta$, $\varphi=\pi/2$, $V=2\Delta/e$, 
		$G_1^{\phi}=0$, and $G_2^{\phi}=2G_N$. (b) Modulus of the spin-singlet, $F_{S}$, and the spin-triplet pairing functions $F_{T0}$, $F_{T\uparrow}$, $F_{T\downarrow}$. Notice, the finite spin-mixing conductance $G^\phi_2$ induces
		a net charge current $I_F$ (top panel) following the curve progression of the equal-spin triplet correlations, thus,
		making them experimentally attainable by standard current measurements [see Fig.~\ref{fig.:sketch}(c) 
		for $\theta\approx0.3\pi$, where $\abs{F_{T\downarrow}}<\abs{F_{T\uparrow}}$].
		}%
	\end{figure}%
	\paragraph{Scaling}
	Before analyzing the interplay between Cooper-pair and quasiparticle transport, let us recall that in equilibrium only a Josephson current $I_S^{\rm eq}=I_{C}\sin\varphi$ 
	may flow between both s-wave superconductors, which has a purely sinusoidal CPR---all other currents require quasiparticle excitations. Depending on the 
	effective size $L$ of the central node, the Josephson current scales in the diffusive regime 
	for $\epsilon_{\rm Th}\ll\Delta$ (large-island) with the Thouless energy $\epsilon_{\rm Th}\equiv\hbar D/L^2$, where $D$ is the diffusion constant [see black solid line in Fig.~\ref{fig.:largeIslandLimit}(a)].
	For $\epsilon_{\rm Th}\gg\Delta$ (small-island), however, it is characterized by the superconducting gap $\Delta$~\cite{Beenakker1992a}.
	While the superconducting condensate is in equilibrium entirely formed by spin-singlet Cooper-pairs, finite voltages may cause triplet correlations in the system. They can lead
	for voltages below the gap [dashed line in Fig.~\ref{fig.:largeIslandLimit}(a)] to a reduction of the Josephson current---here, we consider parallel collinear magnetization. 
	For voltages above the gap and intermediate values of the Thouless energy $\epsilon_{\rm Th}$, such triplet correlations can even induce current reversals in the Josephson current $I_S$. 
	In Fig.~\ref{fig.:largeIslandLimit}(a), showing the modulus of $I_S$ on a double logarithmic scale, these zero-crossings (at which the logarithm diverges) result in the two sharp dips of the dash-dotted curve. Also the corresponding critical current $I_C=\max_\varphi I_S$ [purple line  in Fig.~\ref{fig.:largeIslandLimit}(a)] indicates with the kinks the presence of triplet correlations.
	
	For Thouless energies much smaller (larger) than the superconducting gap, the corresponding net supercurrent stays always positive and is dominated by spin-singlet correlations. 
	We show in the following that applied voltages in combination with non-parallel magnetization, $\theta\neq0$, can induce triplet correlations and transitions in the CPR
	\textit{also} in the regimes $\epsilon_{\rm Th}\ll\Delta$ and $\epsilon_{\rm Th}\gg\Delta$.

	\paragraph{Phase transitions and spin current control}
	First, let us consider the large-island regime, $\epsilon_{\rm Th} \ll \Delta$, where the Cooper-pair transport is characterized by the Thouless energy, i.e. $I_S, I_S^{z}\propto \epsilon_{\rm Th}$.
	While the net current between both superconductors follows in equilibrium the usual sinusoidal CPR, $I_S\propto\sin\varphi$, nonequilibrium in combination
	with non-collinear magnetization, $0<\theta<\pi$, can induce $0$--$\pi$-transitions for voltages $V>0$ below the gap $\Delta$, due to mixed-spin triplet correlations [solid black line in Fig.~\ref{fig.:largeIslandLimit}(b)]. Above the gap, quasiparticle transport sets in and $I_S$ saturates. In this regime, a finite net spin current $I_F^z\propto V$ (red dot-dashed line) emerges between the ferromagnetic contacts for a nonzero magnetization angle, irrespective of $\varphi$. For the chosen symmetric configuration, however, no corresponding net charge current $I_F$ flows.
	
	A special feature of our setup is the occurrence of a finite net spin current $I_S^z$ between both superconductors (blue dashed line) for voltages $V\gtrsim\Delta/e$.
	This effect is maximal for parallel magnetization, $\theta=0$, as can be seen in the inset, for which \textit{only} mixed-triplet and singlet correlations
	are present. It vanishes for antiparallel orientation, $\theta=\pi$, where \textit{no} triplet correlations arise, and when the Josephson phase $\varphi$ is a multiple of $\pi$.
	Notice that $I_S^z$ is antisymmetric in $\varphi$, and $I_S^z$ as well as $I_F^z$ are symmetric in $\theta$. 
	While voltages $V$ above the gap can trigger net spin currents $I_F^z$ and $I_S^z$, the magnetization angle $\theta$ can control their ratio (see inset),
	making the proposed setup, thus, attractive for future applications.

	\paragraph{Spin-mixing induced charge current}
	Let us now turn to the small-island regime, $\epsilon_{\rm Th}\gg\Delta$, where losses of superconducting coherences become irrelevant, and the Josephson transport is characterized by the gap energy, i.e. $I_S,I_S^{z}\propto \Delta$. Here, we find that non-antiparallel magnetization $\theta\ne\pi$ can, indeed, induce triplet correlations for sufficiently large voltages, which results in an asymmetric sinusoidal CPR. However, it \textit{cannot} induce $0$--$\pi$-transitions. To cure this circumstance, we consider in the following finite spin-mixing for voltages above the gap. 
	Indeed, figure~\ref{fig.:finiteSpinMixing}(a) indicates that the system is for parallel magnetization, $\theta=0$, but finite spin-mixing $G_2^\phi>0$ 
	in a $\pi$-phase featuring a negative net supercurrent $I_S$ (black solid line). Roughly at $\theta=\pi/4$, the CPR undergoes a $\pi$--$0$-transition. Similar to
	the large-island regime [inset of Fig.~\ref{fig.:largeIslandLimit}(b)],  a net spin current $I_F^z$ may flow between the ferromagnets 
	[red dot-dashed line in Fig.~\ref{fig.:finiteSpinMixing}(a)] which is essentially unaffected by spin-mixing. The net spin current $I_S^z$ 
	between the superconductors (dashed blue line), on the contrary, is apart from $\theta=0$ modified by spin-mixing, and features a current reversal about $\theta=\pi/2$.
	
	Remarkable, the non-collinear magnetization in this system gives rise to equal-spin triplet correlations $\abs{F_{T\uparrow}}$ and $\abs{F_{T\downarrow}}$, see Fig.~\ref{fig.:finiteSpinMixing}(b). A distinctive feature, however, is that these equal-spin triplet correlations can induce
	a finite net charge current $I_F$ into the ferromagnets [dotted green line in Fig.~\ref{fig.:finiteSpinMixing}(a)] for asymmetric spin-mixing, $G_1^\phi\neq G_2^\phi$. 
	Where, this feature is attributed to the creation of an imbalance in the ferromagnetic spin channels, see Fig.~\ref{fig.:sketch}(c).
	This effect also persists for vanishing Josephson phase $\varphi$. Under a mutual exchange of the spin-mixing conductances ($G_1^\phi\leftrightarrow G_2^\phi$), the net charge current $I_F$ just inverts. 
	An experimentally measurable charge current $I_F$ serves also in the large-island regime as a signature of equal-spin triplet correlations.
	It features in this regime a similar curve progression, but scales instead with $\epsilon_{\rm Th}$.

	\paragraph{Conclusions}
	Spin-dependent quasiparticle and Cooper-pair transport have been analyzed in a proximity-coupled multi-terminal S/F-heterostructure in nonequilibrium. We have shown that $0$--$\pi$-transitions can be induced in the 
	CPR by biasing the ferromagnetic contacts and bearing non-collinear magnetic moments, as long as the loss of superconducting coherences is large, $\epsilon_{\rm Th}\ll\Delta$. 
	In this limit, voltages exceeding the superconducting gap, $V\gtrsim\Delta/e$, trigger net spin currents into the ferromagnets/superconductors, which can be controlled by the relative magnetization angle $\theta$. The small-island regime, however, requires additionally finite spin-mixing
	to induce CPR. The considered heterostructure qualifies as an ideal platform for the generation of triplet correlations of different spin projection, and as a voltage- and phase-controlled switch for spin and electron currents. In particular, it constitutes a minimal setup for generating, controlling and detecting equal-spin triplet correlations by current measurements.

	This work was financially supported by the DFG through SFB 767 and SPP 1538 ``SpinCaT", the Carl-Zeiss-Stiftung, the Alexander von Humboldt Foundation, and the Research Council of Norway through its Centers of Excellence funding scheme, project 262633, ``QuSpin''. A.R. gratefully acknowledges the hospitality of QuSpin (NTNU), during his visit to Trondheim.

\end{document}